\documentclass[
	aps,
    prl,
    twocolumn,
    letterpaper,
    10pt,
    superscriptaddress,
    showpacs,
    notitlepage,
    amsmath,
    amssymb,
    floatfix
]{revtex4}
\pdfoutput=1
\usepackage{bm,enumerate,dcolumn,tikz,graphicx,color,amsmath,amssymb}
\usepackage{epstopdf}
\usepackage{empheq}
\usepackage{capt-of}
\usepackage{pgfplots}
\usepackage{hyperref}% add hypertext capabilities

\newcommand{\ev}[1]{\langle #1 \rangle}
\newcommand{\ket}[1]{| #1 \rangle}
\newcommand{\bra}[1]{\langle #1 |}

\newcommand{\+}{^\dagger}

\newcommand{\bel}{\begin{equation}}
\newcommand{\eel}{\end{equation}}
\newcommand{\be}{\begin{equation*}}
\newcommand{\ee}{\end{equation*}}
\newcommand{\bal}{\begin{eqnarray} }
\newcommand{\eal}{\end{eqnarray}}
\newcommand{\ba}{\begin{eqnarray*}}
\newcommand{\ea}{\end{eqnarray*}}

\newcommand{\reffig}[1]{Fig.~\ref{#1}}

\newcommand{\PP}{\mathcal{P}}

\newcommand{\br}{{\mathbf r}}

\renewcommand{\d}[1]{\! d#1\,}

\renewcommand{\refeq}[1]{Eq.~(\ref{#1})}

\begin{document}
\title{Effects of molecular resonances on Rydberg blockade}
\author{Andrei Derevianko}
\affiliation{Department of Physics, University of Nevada, Reno, NV 89557, USA}
\affiliation{ITAMP, Harvard-Smithsonian Center for Astrophysics, Cambridge, MA 02138, USA }
\author{P\'{e}ter K\'{o}m\'{a}r}
\affiliation{Department of Physics, Harvard University, Cambridge, MA 02138, USA}
\author{Turker Topcu}
\affiliation{Department of Physics, University of Nevada, Reno, NV 89557, USA}
\affiliation{ITAMP, Harvard-Smithsonian Center for Astrophysics, Cambridge, MA 02138, USA }
\author{Ronen M. Kroeze}
\affiliation{Department of Physics, Eindhoven University of Technology, 5600 MB Eindhoven, the Netherlands}
\affiliation{Department of Physics, Harvard University, Cambridge, MA 02138, USA}
\author{Mikhail D. Lukin}
\affiliation{Department of Physics, Harvard University, Cambridge, MA 02138, USA}

\date{\today}

\begin{abstract}
We study the effect of resonances associated with
 complex molecular interaction of Rydberg atoms on Rydberg blockade.
We show that  densely-spaced molecular potentials between doubly-excited atomic pairs  become
unavoidably resonant with the optical excitation at short interatomic
separations.
Such molecular resonances limit the coherent control of individual excitations
in Rydberg blockade.
As an illustration, we compute the molecular interaction potentials of Rb atoms
near the $100s$ states asymptote to characterize such detrimental molecular
resonances,  determine the resonant loss rate to molecules and inhomogeneous
light shifts.  Techniques to avoid the undesired effect of molecular resonances
are discussed.
\end{abstract}

% pacs, the Physics and Astronomy Classification Scheme

\maketitle

%\section{Problem setup}
Rydberg blockade~\cite{LukFleCot01,Jaksch2000,LukinColloquium03,SafWalMo10}
has recently emerged as a promising method for creating and
manipulating quantum states of light and matter in  applications ranging from
quantum information processing~\cite{Isenhower2010,GaeMirWil09,KublerH.2010} to quantum nonlinear
optics~\cite{Hofmann2013,Firstenberg2013}. The key idea is that strong interaction between
Rydberg atoms can be used, under certain conditions, to block the states with
more than one excited atoms.  Multiple Rydberg excitations are suppressed due to
level shifts caused by strong long-range interactions between Rydberg atoms.
This mechanism enables performing quantum logic operations between atom pairs
and manipulate collective  many-body states of $N$-atom
ensemble\cite{LukFleCot01}.
Such collective states efficiently couple to  laser fields with the coupling enhanced by a factor of $\sqrt{N}$, see experiments~\cite{Dudin2012b,EbeKwoWal15}.   While a number of advanced protocols involving Rydberg blockade is  being explored, an outstanding challenge is to identify and realize conditions for high-fidelity atomic and optical state control via Rydberg blockade.

Here we investigate the effect of molecular resonance on quantum state manipulation via Rydberg blockade.
 We demonstrate that the very same interactions that cause the level shifts required for blockade also have  detrimental effects due to a large  state density (number of levels per energy interval) of Rydberg states resulting in a plethora of closely-spaced molecular potentials. Some of these potentials may become, at specific interatomic separation,  resonant with the driving field causing excitations to unwanted doubly-excited Rydberg states. While this mechanism was qualitatively pointed out ~\cite{Keating2013,Dumin2013}, detailed understanding of effects of molecular resonances on collective state manipulation is important for high fidelity quantum states control. This is challenging  partially due to the overwhelming
complexity of molecular potentials especially at small internuclear separations~\cite{Keating2013}. Below we demonstrate that the cumulative effect on the Rydberg blockade is caused by molecular resonances  at large interatomic distances where reliable theoretical predictions can be made. We derive and compute the rates of resonant conversions to diatoms, show that collective qubit rotations are damped, and compute the ``leakage'' and inhomogenous frequency shifts r due to diatom conversion. Finally, we discuss techniques to suppress the deleterious  molecular resonance effects.

\begin{figure}[h!tb]
	\begin{center}
		\includegraphics[width=0.45\textwidth]{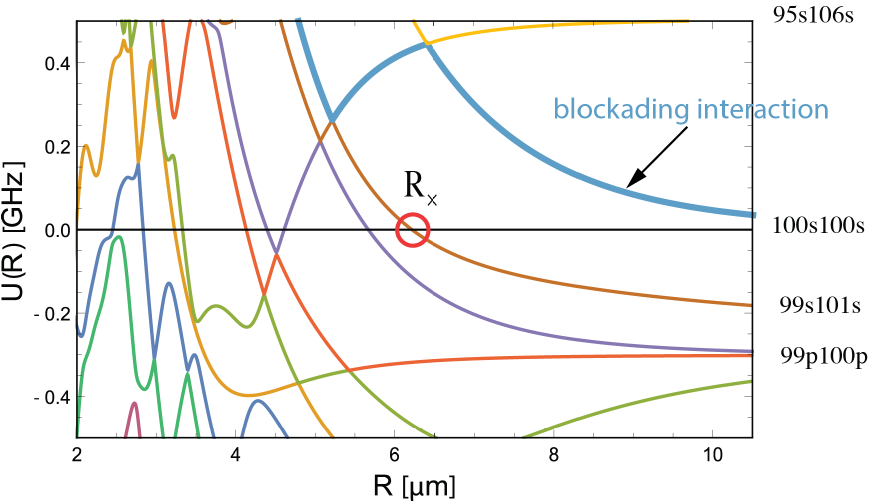}
  \end{center}
\caption{  Selected $\Sigma_g$ molecular potentials   in the $1\, \mathrm{GHz}$ window centered about the $100s +100s$ dissociation limit (placed at zero energy) for two Rb atoms.
The potentials are marked by their double-atom dissociation limits at large internuclear separations $R$.  Highlighted blockading interaction is the interaction that tunes a pair of $100s$ Ry atoms away from
the resonance with driving laser field. The position $R_\times$  of the outer-most resonant molecular potential crossing is marked with a circle. Properties of molecular resonances  are compiled in Table~\ref{Tab:xings}.
	}
	\label{Fig:X1GHzWindow}
\end{figure}

We start by computing molecular potentials for two interacting  Rb Rydberg atoms by a direct diagonalization of the long-range dipole-dipole molecular Hamiltonian.
On a large energy scale, we find a ``spaghetti'' of densely packed curves exhibiting intricate avoided crossing patterns. The region that is relevant to our discussion is centered around the nominally blockaded Rydberg levels. As an illustration, we take $|r\rangle = |100s\rangle$. Considering that the typical excitation Rabi frequency $\Omega_0$ is $\sim 1\, \mathrm{MHz}$ we zoom onto a 1 GHz window (Fig.~\ref{Fig:X1GHzWindow}) centered about the $100s +100s$ dissociation limit. In this figure the potential  that at large $R$ asymptotes  to two $100s$ atoms is the blockading van der Waals interaction.
However, we also find several potential curves that at short $R$ cross zero energy corresponding to a resonance with the laser field.
 As a result, atoms can be promoted into an undesired molecular  state corresponding to two Rydberg atoms.  Properties
 of these resonant crossings are compiled in Table~\ref{Tab:xings}.
  The outermost  crossing  with the most substantial laser coupling is at $R_\times \approx 6.2 \,\mu\mathrm{m}$.  Since this value is larger than the average interatomic
separation for  typical experimental number densities~\cite{Dudin2012b} $\rho_d \approx 10^{11}-10^{12}  \,   \mathrm{cm}^{-3}$, one may find a
fraction of atomic pairs inside the volume enclosing molecular resonance region.  Note that the potential curves were computed in the basis of atomic orbitals with orbital angular momenta up to $\ell_\mathrm{max}=2$.  Increasing $\ell_\mathrm{max}$ and adding  atomic orbitals to the computational basis breed new resonant crossings, as the system becomes increasingly chaotic at smaller $R$ due to stronger inter-channel couplings and thus larger number of avoided crossings. Even our outermost resonance can be superseded by
crossings at larger $R$, but with  suppressed laser couplings.
 However, the parameters of the outermost crossing in Fig.~\ref{Fig:X1GHzWindow} are stable with respect to the basis variation. As we demonstrate below namely this outermost crossing predominantly affects
 the dynamics of collective excitations thereby mitigating challenges
 of reliably computing  full-scale Ry-Ry interaction potentials.

\begin{table}[h]
\centering
\begin{tabular}{dddd}
\hline \hline
\multicolumn{1}{c}{$R_\times, \mu\mathrm{m}$} &
\multicolumn{1}{c}{$\xi_m$} &
\multicolumn{1}{r}{ $\Delta  R_\times, \mu\mathrm{m}$ } &
\multicolumn{1}{r}{$\gamma_m, \mathrm{s}^{-1}$} \\
%$R_\times, \mu\mathrm{m}$ & $\xi_m $ & $\Delta  R_\times, \mu\mathrm{m}$ & $\gamma_m, \mathrm{s}^{-1}$ \\
%R_\times, \mu\mathrm{m} & \xi_m = \Omega_m/\Omega_0 & \Delta  R_\times, \mu\mathrm{m} & \gamma_m, \mathrm{s}^{-1} \\
\hline\\[-1ex]
%1
6.22 & 0.55 & 3.8\times 10^{-4} & 1.0\times 10^5\\
%2
5.67 & 0.091 & 3.5\times 10^{-5} & 1.3\times 10^3\\
%3
4.61 & 0.012 & 2.\times 10^{-6} & 6.2\times 10^0\\
%4
4.39 & 0.44 & 8.6\times 10^{-5} & 9.1\times 10^3\\
%5
4.13 & 0.13 & 1.7\times 10^{-5} & 4.6\times 10^2\\
%6
3.33 & 0.16 & 9.3\times 10^{-6} & 2.0\times 10^2\\
%7
2.45 & 0.011 & 1.2\times 10^{-6} & 9.8\times 10^{-1}\\
%8
1.99 & 0.0017 & 6.4\times 10^{-8} & 5.4\times 10^{-3}\\
\multicolumn{4}{c}{$\cdots$}\\
\hline\hline
\end{tabular}
\caption{\label{Tab:xings} Molecular resonance shell properties for the Rb $100s$ blockaded  state (see Fig.~\ref{Fig:X1GHzWindow}). $R_\times$ and $\Delta R_\times$ are the shell radii and widths, $\xi_m$ are the fractional molecular Rabi frequencies, $\xi_m = \Omega_m/\Omega_0$, and $\gamma_m$ are molecular loss rates. $\Delta R_\times$ and $\gamma_m$ are evaluated for atomic Rabi frequency $\Omega_0= (2 \pi) \times 0.1 \, \mathrm{MHz}$ and number density $\rho_d = 10^{12} \, \mathrm{cm}^{-3}$. Positions and the number of resonances for smaller $R$  are computational-basis dependent, however,  their contribution to the total rate is strongly suppressed.  }
\end{table}

Despite the complexity of molecular potentials, the position $R_\times$ of the outermost resonance can be  estimated as follows. Suppose the $ns+ns$ state is our  $|rr\rangle$ ``blockaded'' state. The nearest-energy $n's+n''s$ state with $n'\approx n'' \approx n$ is  the $(n-1)s + (n+1)s$ state and at large $R$ it lies below the resonance by $\delta_m \approx-3n^{-4}$.
Further, the molecular potential correlating to the  $(n-1)s + (n+1)s$  atoms, behaves as $n^4 \tilde{c}_3/R^3$ due to the repulsion from the $p+p$ state below, where $ \tilde{c}_3 \sim 1$. Thereby, $U(R) \approx -3n^{-4} + n^4 \tilde{c}_3/R^3$ and from $U(R_\times)=0$ we arrive at   $R_\times \approx (3^{-1} \tilde{c}_3 \,  n^8)^{1/3}$. For $n=100$ this estimate with  $\tilde{c}_3 = 1$ leads to $R_\times \approx 8 \,\mu\mathrm{m}$ in a reasonable agreement with our computed value.
\begin{figure}[h!tb]
	\begin{center}
		\includegraphics[width=0.4\textwidth]{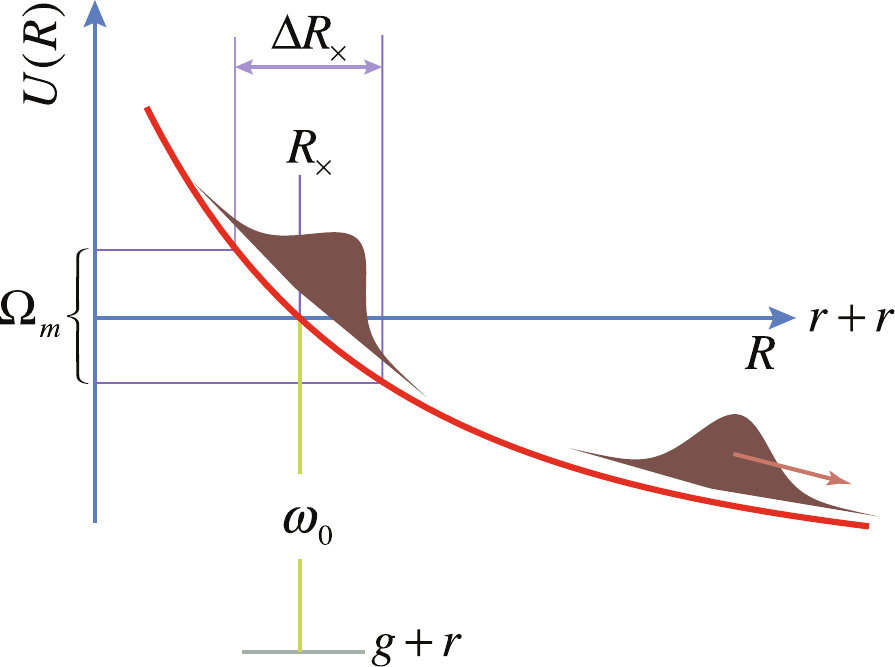}
  \end{center}
\caption{  A laser pulse nominally resonant with the $r-g$ transition can resonantly excite molecular states. The molecular wave-packet is efficiently excited within a window $\Delta R_\times$ determined by the slope of molecular potential $U(R)$ and molecular Rabi frequency $\Omega_m$. Once excited the wave-packet rapidly accelerates out of resonance and rolls down
the slope of molecular potential.
	}
	\label{Fig:resonance-shell}
\end{figure}
Further, we evaluated molecular Rabi frequencies $\Omega_m=\xi_m \Omega_0$
(typically a fraction $\xi_m$ of $\Omega_0$, see  Table~\ref{Tab:xings}).
%For the outermost resonance, such couplings originate from admixtures of the
%$100s+ns$ states through the off-diagonal van der Waals interaction.
%We
%evaluated  $\xi_m$
from the eigenvectors of the numerically diagonalized
molecular Hamiltonian.

%To estimate the effect of these resonances, we note that
The atoms are
efficiently laser-coupled to the molecular  resonances only in a small window of $R$, when the detuning $U(R)$ is comparable to $\Omega_m$ (see Fig.~\ref{Fig:resonance-shell}). Thereby we  define an effective radial width of the molecular resonance
\begin{equation}
 \Delta R_\times = \Omega_m(R_\times) / |U'(R_\times)|  \, , \label{Eq:dRx}
 \end{equation}
 where $\Omega_m$ and the derivative of the molecular potential $U'(R_\times) = dU(R_\times)/dR$ are evaluated at the resonance crossing. For the outermost resonance $\Delta R_\times \approx \Omega_0 \xi_m n^{20/3} (\tilde{c_3}/3)^{1/3} /9$.   Each molecular resonance therefore defines a ``resonance shell'', a spherical shell of radius $R_\times$ and radial width $\Delta R_\times$ centered at a given atom.
The average number of atoms inside the resonance shell,
\begin{equation}
\Delta N_\times = 4 \pi R_\times^2 \Delta R_\times \rho_d \, ,
\end{equation}
 is a relatively small number in a typical experiment. For parameters of Table~\ref{Tab:xings},  the outermost resonance shell contains less than $\sim$ 0.1 atoms.

The presence of molecular resonances implies several consequences for the Rydberg blockade, the most important being the atomic loss.
Indeed, inside the shell two excited Rydberg atoms are a subject to a mechanical force $-U'(R_\times)$. This force can be either attractive or repulsive. (An example of the repulsive resonance is at $R_\times \approx 6.2 \, \mu\mathrm{m}$, see Fig.~\ref{Fig:resonance-shell}). The diatom would separate
into two $99s$ and $101s$ Ry atoms with a kinetic energy of relative motion equal to the dissociation limit  $\delta_m \approx-3n^{-4}$, which is  $\sim 10 \,\mathrm{mK}$ for $n=100$.  These atoms may escape the trapping volume, effectively reducing the number of blockaded atoms. Since the atoms are accelerated out of the resonance shell on timescales $\tau_a = \sqrt{\Delta R_\times} M_a/|U'(R_\times)| \ll 1/\Omega_0$,
we adopt  a simple model that once a pair of atoms is promoted to a molecule, the associated motional wave packet quickly leaves the resonance shell with its atomic constituents no longer  interacting with the laser field.

Now we analyze the dynamics of collective atomic ensemble excitations. We will derive the expressions for the damping (loss) rate in two approximations:  (i) assuming that atoms are frozen in space (static limit) and  (ii) collisional model (impact limit). Remarkably both approximations  yield identical loss rate. We consider an ensemble  of $N$ atoms initially in the collective ground state $|G\rangle = |g_1g_2\cdots g_N\rangle$. A laser pulse couples $|G\rangle$  to a superposition of singly excited Rydberg atoms $|R_i\rangle = |g_1\cdots g_{i-1}r_ig_{i+1} \cdots g_N \rangle$.
These atoms can be further promoted to doubly-Ry-excited diatom states $|M_{ij}\rangle$, involving atoms $i$ and $j$. There are $N_m=N(N-1)/2$ diatom states, with their resonance detunings $\Delta(R_{ij})=U(R_{ij})$ and Rabi frequencies  $\Omega_m^{ij}(R_{ij})$ determined by their interatomic separations $R_{ij}$.  Expanding the total wavefunction in this basis ($ \omega_0$ is the laser frequency  resonant with the $g-r$ transition)
\[
|\Psi \rangle = c_g e^{i\omega_0 t} |G\rangle + \sum_i^N c_i    |R_i\rangle +  e^{-i\omega_0 t}  \sum_i^N \sum_{j>i}^N m_{ij}  |M_{ij}\rangle
\]
and applying the rotating-wave approximation, we arrive at
\begin{eqnarray*}
i \dot{c}_g &=& \frac{1}{2} \Omega_0 \sum_i^N c_i  \,, \\
i \dot{c}_i &=&  \frac{1}{2}\Omega_0 c_g +  \frac{1}{2}\left( \sum_{j=1}^{i-1}  \Omega_m^{ij}  m_{ji} + \sum_{j=i+1}^N  \Omega_m^{ij}  m_{ij} \right)     \,, \\
i \dot{m}_{ij} &=&  \Delta_{ij} m_{ij}+  \frac{1}{2} \Omega_m^{ij}  \left( c_i +c_j \right)    \,.
\end{eqnarray*}
When all molecular detunings  $\Delta_{ij}$ are large, the system undergoes the ideal Rabi flopping between the collective ground state and symmetric combination of single Rydberg excitations ($c^I_g(t) = \cos(\sqrt{N} \Omega_0 t/2)$ and  $c_i^I(t) =- i\sin( \sqrt{N} \Omega_0 t/2)/\sqrt{N}$). We focus on the averaged collective dynamics  and introduce collective amplitude $c_s$ so that $c_i(t) \equiv c_s(t)/\sqrt{N}$. We assume that all $\Omega_m^{ij}=\Omega_m$ owing to the weak $R_{ij}$ dependence inside the resonance shell. The collective amplitudes satisfy ($\Omega_0^N = \sqrt{N} \Omega_0$) \begin{align}
i \dot{c}_g &= \frac{ \Omega_0^N}{2}  c_s \,, \nonumber \\
i \dot{c}_s &=  \frac{\Omega_0^N}{2} c_g + \frac{\Omega_m}{2\sqrt{N} }\sum_i^N \left( \sum_{j=1}^{i-1}   m_{ji} + \sum_{j=i+1}^N  m_{ij} \right)    \nonumber \,, \\
i \dot{m}_{ij} &=  \Delta_{ij} (R_{ij}) m_{ij}+   (\Omega_m/\sqrt{N})  c_s   \label{Eq:mdotCollective} \,.
\end{align}
%{\em Static approximation---}

Now we fix the positions of all atoms (this requirement is relaxed later on) and
split the time axis into time intervals consisting of a short laser pulse of duration $\tau_p \ll 1/\Omega_0^N$ and field-free acceleration time $\tau_a$ during which
the excited diatom wave packet leaves the shell. Because of the mechanical
forces the molecular amplitudes inside the shell are reset to zero values before the next pulse arrives (this is reminiscent of the Markov approximation~\cite{blum2012density}). By taking the limit $\tau_a \rightarrow 0$ we arrive at a continuous Rabi drive.  Integrating the last equation over time interval $(t,t+\tau_p)$, one obtains,
$m_{ij}(t+\tau_p) =
 c_s(t) (\Omega_m/\sqrt{N})
\left\{ \exp(i \Delta_{ij}  \tau_p) -1\right\}/\Delta_{ij}$.
We have set $m_{ij}(t)=0$ as discussed. Notice that the r.h.s. spikes at $\Delta_{ij}=0$, i.e., within the resonance shell. Ensemble averaging yields
 $\langle m_{ij}(t+\delta t) \rangle = -i c_s(t) \pi (\Omega_m/\sqrt{N} )(4 \pi R_\times^2)/(|U'(R_\times)| V_s)$, where $V_s$ is the blockaded ensemble volume. By substituting this relation into the equation for $c_s$ we
arrive at a set of damped equations (non-Hermitian Schrodinger equation)
 \begin{align}
i \dot{c}_g &= \frac{ \Omega_0^N}{2}  c_s \,, \nonumber\\
i \dot{c}_s &=  \frac{\Omega_0^N}{2} c_g - i \gamma_m c_s \, \label{Eq:damped}
\end{align}
with the molecular-resonance (amplitude) loss rate
\begin{equation}
 \gamma_m = \pi \Delta N_\times \Omega_m/2 = 2 \pi^2 \rho_d \xi_m^2  R_\times^2 \Omega_0^2/|U'(R_\times)| \,.
 \label{Eq:rate}
\end{equation}
For the outermost resonance,  $\gamma_m  \approx 2\pi^2 \rho_d \xi_m^2 \Omega_0^2  \tilde{c}_3\, n^{12} /27$.
The above derivation neglected atomic motion and is valid for very cold
ensembles. In the SM, we take into account the thermal
motion of the atoms, using the impact approximation. We find the result for
$\gamma_m$ that is identical to Eq.~(\ref{Eq:rate}).

In addition to the loss, the same $\Omega$ drive induces an AC Stark shift which
is different for states in which different atoms are excited. This inhomogeneous
broadening results in an additional loss of coherence of the Rabi oscillation.
In the SM, we show that this broadening is given by
\begin{equation}
	\delta\Delta \approx (7\pi/2) \rho_d \xi_m^2 R_\times^2 \Omega_0^2 /
	|U'(R_\times)|
\end{equation}
in the limit where $R_\times$ is much smaller than the size of the ensemble.
Although the dependence of the system parameters is the same as for $\gamma_m$
in Eq.(\ref{Eq:rate}), the prefactor makes $\delta\Delta$ roughly two times
smaller than $\gamma_m$. During the Rabi dynamics, the  broadening makes the signal decay as $\exp[-(t\delta\Delta)^2]$ over
time.

The rate formula is to be summed over all
resonance shells:
$\gamma_m^\mathrm{tot} = \sum_k \gamma_m^k$, where $\gamma_m^k$ is the individual shell contribution~(\ref{Eq:rate}). For our example in Table~\ref{Tab:xings}, $\gamma_m^\mathrm{tot}$ is entirely dominated by the
 outermost crossing. The reason
 for this prominence is that at smaller $R$, the potentials become steeper and the molecular Rabi frequencies become diluted.
% , thereby leading to smaller values of $\Delta R_\times$ (see Eq.(\ref{Eq:dRx}) and Table~\ref{Tab:xings}) and together with smaller values of $R_\times$ leading to smaller values of $\Delta N_\times$ and thereby $\gamma_m$.
 Notice that the long-range molecular Hamiltonian used in computing the molecular potential curves in Fig.~\ref{Fig:X1GHzWindow} holds only for $R \gg 2 n^2 a_0 \sim 1\, \mu m$ for $n=100$, i.e. when the electronic densities do not overlap. However, all the qualitative arguments that the molecular excitation rates should be suppressed compared to the outermost resonance shell still hold even for small $R$.

The rate scales steeply with $n$, $\gamma_m \propto n^{12}$.
In particular, it is commonly believed that the blockade fidelity can be improved by going to high-$n$ Ry states, because the probability of off-resonant Ry excitations is suppressed as $n^{-22}$ in the van der Waals blockade.
We see that increasing $n$ while suppressing off-resonant Ry excitations also increases the undesired molecular loss rates.

%{\em Damped oscillations---}
 Eqs.~(\ref{Eq:damped}) reduce to the damped oscillator equation of motion $\ddot{c}_s + (\Omega_0^N/2)^2 c_s - \gamma_m c_s=0$  with  solutions
\begin{align}
c_s(t) &=  -i  \left(\Omega_0^N/\Omega_d\right) \sin\left( \Omega_d t/2\right) e^{-\frac{\gamma_m}{2}t}\,,\\
c_g(t) &=  \left( \cos\left( \Omega_d t/2\right)  + (\gamma_m/\Omega_d) \sin\left( \Omega_d t/2\right) \right)e^{-\frac{\gamma_m}{2}t} \, , \label{Eq:DampedSolutions}
\end{align}
where $\Omega_d = \Omega_0^N \left( 1 - \eta^2 \right)^{1/2}$,  $\eta = \gamma_m/\Omega_0^N$. Driven  ensemble exhibits
 damped collective Rabi oscillations with a frequency $\Omega_d \le \Omega_0^N$. One may distinguish between three classes of solutions~\cite{Georgi2007book}: under-damped ($\eta < 1$), critical ($\eta=0$) and over-damped ($\eta > 1$). Explicitly,
 \[
 \eta = 2 \pi^2 \frac{ \sqrt{N}}{V_s |U^\prime(R_\times)|} \xi_m^2 R_\times^2 \Omega_0
 \propto n^{12} \sqrt{N} \Omega_0.
 \]
 Thereby increasing $\Omega_0$, $n$, or $\rho_d$
  can cause the ensemble to exhibit over-damping of collective Rabi oscillations,
  at which point they no longer resemble oscillations. In the under-damped
  regime, the loss per collective Rabi cycle determines collective qubit operation fidelity
  $F = (\Omega_0^N/\Omega_d)^2 e^{-2\pi\eta}$.

The molecular loss  can account for some experimentally observed imperfections.
E.g., Dudin et al.
\cite{Dudin2012b} have effectively measured  the damping constant for collective
Rabi oscillations in a mesoscopic ensemble of $102s$ Rb atoms. They found that
the Rabi oscillation loses $10-20\%$ of its contrast in a single cycle. Our
calculation can account for a loss of $\sim 5\%$ during a single oscillation
cycle. While the agreement seems to be adequate, we  emphasize that it may be
fortuitous as the experiment has been carried out in the presence of magnetic
field  which was excluded in our analysis and would introduce additional
resonances. We also neglected the $s-d$ excitation channels (allowed in the
excitation scheme~\cite{Dudin2012b}) when computing molecular Rabi
frequencies. Moreover, the experiment~\cite{Dudin2012b} is affected by a
multitude of other decoherence effects.
At this point it may be desirable to design experiments that could disentangle
various decoherence mechanisms, and the molecular losses in particular.

While the total number of atoms $N$ remains constant during the coherent  evolution, the wave function acquires  out-coupled diatom wave packets.
If the measurement of the total number of atoms were to be made, the number of atoms remaining in the ensemble would be $N_\mathrm{eff}(t)=N (|c_g(t)|^2 + |c_s(t)|^2)$. By manipulating Eq.~(\ref{Eq:damped}), one finds that $\dot{N}_\mathrm{eff} = - 2 \gamma_m |c_s(t)|^2 N$, resulting in  $
 N_\mathrm{eff}(t) = N \left\{
 1 + 2  \left(\frac{\gamma_m}{\Omega_d} \right)^2 \sin^2(\frac{\Omega_d t}{2}) +
  \left(\frac{\gamma_m}{\Omega_d} \right) \sin(\Omega_d t)
 \right\} e^{ - \gamma_m t}
$, or averaging over many cycles $\bar{N}_\mathrm{eff}(t) = N (  1 +  \left(\gamma_m/\Omega_d \right)^2)e^{ - \gamma_m t}$, i.e., the effective number of atoms remaining in the ensemble decays exponentially. The quantity $1-(|c_g|^2+ |c_s|^2) = (N- N_\mathrm{eff}(t))/N\sim \left(\gamma_m/\Omega_d \right)^2 e^{ - \gamma_m t}$ also determines ``leakage'' from the collective qubit space. Clearly, to minimize the leakage one has
to require that $\gamma_m t \ll 1$ or $\gamma_m \ll \Omega_0 \sqrt{N}$. For parameters of
Table~\ref{Tab:xings}, the coherent evolution is limited to $t \ll 10 \mu\mathrm{s}$.

%We have developed an understanding of leakage from the collective qubit space
%due to out-coupling of resonant pairs of diatoms.
One may visualize the
``leakage'' from the collective qubit space as a modulated outflow of molecular wave packets from the blockaded
volume. As an illustration, the outermost molecular resonance produces
admixtures of $101s$ and $99s$ Rydberg atoms.
If the ensemble is trapped, the out-coupled (di)atoms may linger inside the
ensemble depending on the released kinetic energy and the trapping
potential height. Such atoms do not resonantly interact with the laser field of the
Rabi drive. However, they do interact with the remaining ensemble leading to energy shifts through the
interactions with the remaining atoms.  Such mechanisms can be also relevant for
untrapped ensembles, where the outflowing diatom wave packets may  interact with
the remaining ensemble while transiting out through the volume. In addition, the present discussion focused on
Rydberg S-states, the undesired effects can be
enhanced for Rydberg states with higher angular momentum. This is due to
the presence of closely spaced states, due to, e.g., spin-orbit interaction, that
can result in molecular crossing at larger $R$.

The unwanted effect of molecular resonances can be suppressed by using tight
traps for individual atoms prior to excitation, such as optical lattices.  The  idea is positioning atoms such that excitation to
molecular resonances is not allowed.  As shown in  the SM,  the loss can be suppressed by a factor of $\sim 100$, if the tightly
trapped  ($\sim 20~\text{nm}$) individual atoms are prepared in a 3D optical
lattice with the lattice constant  tuned to avoid the resonances.  By choosing the lattice constant, the outermost resonant shell $R_\times = 6.2~\mu\text{m}$ can fall in a gap between density peaks,  largely reducing its effect. High fidelity manipulation of Rydberg atoms in a lattice has been observed recently~\cite{Zeiher2015},
% J. Zeiher, P. Schau§, S. Hild, T. Macr, I. Bloch, and C. Gross, arXiv Prepr. 1 (2015).

We investigated how  molecular resonances limit the fidelity of
Rydberg excitations in an atomic cloud. Under
continuous driving pairs of atoms can be promoted into a doubly excited Rydberg
states, if they are separated by certain resonant distances. These resonant pairs
repel each other and leave the cloud.
%This limits the fidelity of  operations relying on dipole blockade.
To mitigate this detrimental effect,  trapping
the atoms in a tight optical lattice can be used, where they are kept away from resonance.

\begin{acknowledgments}
We would like to thank E. Tiesinga and A. Kuzmich for discussions.
This work was supported in part  by the US National Science Foundation (NSF) Grant No. PHY-1212482. A.D. was also supported by the Simons foundation as a Simons fellow in theoretical physics. T.T. and A.D. would like to thank the Institute for Theoretical Atomic, Molecular and Optical Physics (ITAMP), Center for Ultracold Atoms and the Harvard University Physics Department for their hospitality. Work at Harvard was supported by NSF, CUA, AFOSR MURI and NSSEFF program.
\end{acknowledgments}
\newpage
\begin{widetext}
{\bf Supplementary Materials}
\section{Derivation of the molecular excitation rate in the impact approximation}

As discussed in the main text, the collisions leading to strong coupling to
molecular states are short and well-separated. Let's consider one of such
collisions of atom $i$ with an atom $j.$ The molecular probability amplitude
satisfies the equation%

\begin{equation}
i\dot{m}_{ij}=\Delta_{ij}m_{ij}+\frac{1}{2}\Omega_{m}^{ij}\left(  c_{i}%
+c_{j}\right)  .\, \label{Eq:EOMmol}%
\end{equation}
The detunings $\Delta_{ij}$ (molecular potentials with the zero energy at the
$|r\rangle+|r\rangle$ dissociation limit, $\Delta_{ij}=U\left(  R_{ij}\left(
t\right)  \right)  $) and Rabi frequencies are time dependent because of the
atomic motion. Introducing%
\[
m_{ij}\left(  t\right)  =\tilde{m}_{ij}\left(  t\right)  \exp\left(
-i\int_{-\infty}^{t}\Delta_{ij}\left(  t^{\prime}\right)  dt^{\prime}\right)
,
\]
we recast Eq.(\ref{Eq:EOMmol}) into
\begin{equation}
i\frac{d}{dt}\tilde{m}_{ij}\left(  t\right)  =\frac{1}{2}\Omega_{m}%
^{ij}\left(  c_{i}+c_{j}\right)  \exp\left(  i\int_{-\infty}^{t}\Delta
_{ij}\left(  t^{\prime}\right)  dt^{\prime}\right)  \label{Eq:mtildeODE}%
\end{equation}
We are interested in the probability of molecular excitation due to a single
collision,%
\[
P_{m}=\left\vert m_{ij}\left(  \infty\right)  \right\vert ^{2}=\left\vert
\tilde{m}_{ij}\left(  \infty\right)  \right\vert ^{2}.
\]
Integrating Eq.(\ref{Eq:mtildeODE}), we arrive at
\begin{equation}
\tilde{m}_{ij}\left(  \infty\right)  =-i\int_{-\infty}^{\infty}\frac{1}%
{2}\Omega_{m}^{ij}\left(  t\right)  \left(  c_{i}\left(  t\right)
+c_{j}\left(  t\right)  \right)  \exp\left(  i\int_{-\infty}^{t}U\left(
R_{ij}\left(  t^{\prime}\right)  \right)  dt^{\prime}\right)  dt
\label{Eq:OscIntegral}%
\end{equation}
To evaluate this probability we approximate $R_{ij}\left(  t\right)  $ with
straight-line trajectories,%
\[
R_{ij}(t)=\left\{  (v\left(  t-t_{c}\right)  )^{2}+\rho^{2}\right\}  ^{1/2}.
\]
Here $\rho$ is the conventional impact parameter, $v$ is the relative atomic
velocity, and $t_{c}$ is the time of the closest approach. The atoms reach the
resonance region when $R_{ij}(t)=R_{\times}$. Clearly, one has to require that
$\rho\leq R_{\times}$ for this to occur. The associated moment of time
$\ t_{\times}$ is
\[
v\left(  t_{\times}^{\pm}-t_{c}\right)  =\pm\sqrt{R_{\times}^{2}-\rho^{2}}.
\]
The exponential in \ the integral (\ref{Eq:OscIntegral}) rapidly oscillates
except when the phase $\int_{-\infty}^{t}U\left(  R_{ij}\left(  t^{\prime
}\right)  \right)  dt^{\prime}$ is stationary. The prefactor varies slowly in
time compared to the exponent. This forms the basis for evaluating
(\ref{Eq:OscIntegral}) using the stationary-phase method. Let's review the
basics of this method. Consider an integral%
\[
I=\int_{-\infty}^{+\infty}g\left(  t\right)  e^{i\phi\left(  t\right)  }dt,
\]
where $g\left(  t\right)  $ varies slowly compared to the rapidly oscillating
exponent. The main value of the integral is accumulate in the regions where
the phase is stationary, i.e.%
\[
d\phi\left(  t_{\ast}\right)  /dt=0.
\]
Expanding the phase in the vicinity of the $t_{\ast}$%
\[
\phi\left(  t\right)  \approx\phi\left(  t_{\ast}\right)  +\frac{1}{2}%
\phi^{\prime\prime}\left(  t_{\ast}\right)  \left(  t-t_{\ast}\right)  ^{2}.
\]
Then%
\begin{align*}
I  &  \approx g\left(  t_{\ast}\right)  e^{i\phi\left(  t_{\ast}\right)  }%
\int_{-\infty}^{+\infty}\exp\left(  i\frac{1}{2}\phi^{\prime\prime}\left(
t_{\ast}\right)  \left(  t-t_{\ast}\right)  ^{2}\right)  dt=\\
&  g\left(  t_{\ast}\right)  e^{i\phi\left(  t_{\ast}\right)  }\exp\left(
i\frac{\pi}{4}~sign\left(  \phi^{\prime\prime}\left(  t_{\ast}\right)
\right)  \right)  \sqrt{\frac{2\pi}{\left\vert \phi^{\prime\prime}\left(
t_{\ast}\right)  \right\vert }}%
\end{align*}
In our case $\phi\left(  t\right)  =\int_{-\infty}^{t}U\left(  R_{ij}\left(
t^{\prime}\right)  \right)  dt^{\prime}$, and the stationary points correspond
to the crossing of the resonance shell%
\[
d\phi\left(  t_{\ast}\right)  /dt=U\left(  R_{ij}\left(  t_{\ast}\right)
\right)  =0,
\]
i.e. $t_{\ast}=t_{\times}$. Notice that we have two stationary points
corresponding to two crossings of the resonance shell. The two times are
separated by
\[
t_{\times}^{+}-t_{\times}^{-}=\frac{2\sqrt{R_{\times}^{2}-\rho^{2}}}{v}%
\]
In general, both points can contribute. However, once promoted to the
molecular state, the atoms experience strong mechanical forces and are
accelerated out of the resonance. Therefore we will neglect the interference
effects when computing the probability $p_{m}$ and add the two contributions
incoherently (this provides the upper limit on $P_{m}$)
\[
P_{m}=\pi\Omega_{m}^{2}\left\vert c_{i}\left(  t_{x}\right)  +c_{j}\left(
t_{x}\right)  \right\vert ^{2}\left(  \frac{1}{\left\vert \phi^{\prime\prime
}\left(  t_{x}\right)  \right\vert }\right)
\]
Further we evaluate the second derivative of the phase evaluated at crossing
points%
\begin{align*}
d^{2}\phi\left(  t_{x}\right)  /dt^{2}  &  =\frac{dU}{dR_{ij}}\frac{dR_{ij}%
}{dt}=\frac{\Omega_{m}}{\Delta R_{\times}}\frac{v}{R_{\times}}v\left(
t_{\times}-t_{c}\right)  =\\
&  \frac{\Omega_{m}}{\Delta R_{\times}}\frac{v}{R_{\times}}\sqrt{R_{\times
}^{2}-\rho^{2}}%
\end{align*}
From here one could define the effective duration of collision%
\[
\tau_{c}\approx\sqrt{\frac{\Delta R_{\times}}{v}\frac{1}{\Omega_{m}}}%
\]
or%
\[
P_{m}=\pi\left\vert c_{i}\left(  t_{x}\right)  +c_{j}\left(  t_{x}\right)
\right\vert ^{2}\Omega_{m}\frac{\Delta R_{\times}}{v}\frac{R_{\times}}%
{\sqrt{R_{\times}^{2}-\rho^{2}}}%
\]
We further approximate the time evolution of single-Ry-excitations via their
un-coupled time evolution, $c_{k}^{I}(t)=i/\sqrt{N}\sin(\sqrt{N}\Omega
_{0}t/2))$%
\[
P_{m}\left(  \rho,t\right)  =\frac{4\pi}{N}\Omega_{m}\frac{\Delta R_{\times}%
}{v}\frac{R_{\times}}{\sqrt{R_{\times}^{2}-\rho^{2}}}\sin^{2}(\sqrt{N}%
\Omega_{0}t/2)).
\]
$p_{m}=0$ for $\rho>R_{x}$ in the stationary-phase approximation as there are
no crossing through the resonance region for such impact parameters.

Number of atoms lost due to a single collision is $\Delta N=-2P_{m}$. Now we
sum the probabilities over multiple collisions. Number of atoms in a relative
velocity group $dv~$passing through the area $2\pi\rho d\rho$ per time
interval $dt$ is equal to $2\pi n\rho d\rho\left\vert v\right\vert f\left(
v\right)  d^{3}vdt$, where $n$ is the number density and $f\left(  v\right)  $
is the velocity distribution. Then the compound atom loss satisfies the
equation (here the factor of $1/2$ is introduced to correct for
double-counting)%
\begin{align*}
\frac{dN}{dt}  &  =-2\gamma_{m}\left(  t\right)  \frac{N}{2}=-\gamma
_{m}\left(  t\right)  N,\\
\gamma_{m}\left(  t\right)   &  =\int\int2\pi n\rho~d\rho~\left\vert
v\right\vert f\left(  v\right)  d^{3}v~p_{m}%
\end{align*}
Explicit evaluation yields the cross-section%
\[
\sigma_{m}\left(  t\right)  =2\pi\int_{0}^{R_{\times}}\rho~d\rho~p_{m}\left(
\rho,t\right)  =\left(  2\pi~R_{\times}^{2}\right)  ~\frac{4\pi}{N}\Omega
_{m}\frac{\Delta R_{\times}}{v}\sin^{2}(\sqrt{N}\Omega_{0}t/2)).
\]
and the rate%
\[
\gamma_{m}\left(  t\right)  =n\left(  2\pi~R_{\times}^{2}\right)  ~\frac{4\pi
}{N}\Omega_{m}\Delta R_{\times}\sin^{2}(\sqrt{N}\Omega_{0}t/2))
\]
For a spherical volume of radius $R_{s},$ $n=3N/\left(  4\pi R_{s}^{3}\right)
$, thereby%
\[
\gamma_{m}\left(  t\right)  =6\pi~~\left(  \frac{R_{\times}}{R_{s}}\right)
^{3}\left(  \frac{\Delta R_{\times}}{R_{\times}}\right)  \Omega_{m}\sin
^{2}(\sqrt{N}\Omega_{0}t/2))
\]
The rate equation has the solution%
\begin{align*}
N\left(  t\right)   &  =N\left(  0\right)  \exp\left(  -\int_{0}^{t}\gamma
_{m}\left(  t^{\prime}\right)  dt^{\prime}\right)  ,\\
\int_{0}^{t}\gamma_{m}\left(  t^{\prime}\right)  dt^{\prime}  &  =\bar{\gamma
}_{m}\left(  t-\frac{\sin\left(  \sqrt{N}\Omega_{0}t\right)  }{2\sqrt{N}%
\Omega_{0}}\right)  ,\\
\bar{\gamma}_{m}  &  =3\pi~~\left(  \frac{R_{\times}}{R_{s}}\right)
^{3}\left(  \frac{\Delta R_{\times}}{R_{\times}}\right)  \Omega_{m}.
\end{align*}

For sufficiently long time ($t\gg4\pi/\left(  \sqrt{N}\Omega_{0}\right)  $),
the total number of atoms falls off exponentially as%
\[
N\left(  t\right)  =N\left(  0\right)  \exp\left(  -\bar{\gamma}_{m}t\right)
.
\]

Finally, the experiments are carried out with mesoscopic ensembles and as
discussed in the main text, the radius of the resonance shell $R_{\times}$
maybe comparable to $R_{s}$ (or blockade radius). It is clear that if
$R_{\times}>2R_{s}$, the atoms are not going to be affected by that particular
molecular resonance. In analogy with Eq.() of the main text we may further
introduce a geometric probability factor $g\left(  R_{\times}/R_{s}\right)  $.%
\[
\gamma_{m}\left(  t\right)  \rightarrow\gamma_{m}\left(  t\right)  g\left(
R_{\times}/R_{s}\right)
\]

Further rates from multiple resonance add%
\[
\gamma_{m}\left(  t\right)  \rightarrow\sum_{k}\gamma_{m}^{k}\left(  t\right)
\]
where $\gamma_{m}^{k}$ is the rate due an individual resonance shell at
$R_{\times}^{k}$.

\section{Derivation of the inhomogeneous broadening}
\subsection{Hamiltonian}
We assume that the $N$ identical atoms move negligibly over the entire extent of
the dynamics in question, so we need to track only the electronic degrees of
freedom. We model each atom as a four-level system, with states
$\ket{g}$, $\ket{r}, \ket{r'}, \ket{r''}$. Let us define the following
collective states,
\bal
	\ket{G} &=& \ket{g}^{\otimes N},
	\\
	\ket{j} &=& \sigma_j\+ \ket{G},\qquad \text{where}\quad \sigma_j =
	\ket{g}_j\bra{r}_j,\qquad j = 1,2,\ldots N,
	\\
	\ket{j,k} &=& {\sigma'}_j\+ {\sigma''}_k\+ \ket{G},\qquad \text{where}\quad \sigma'_j =
	\ket{g}_j\bra{r'}_j,\quad \text{and}\quad \sigma_j'' = \ket{g}_j\bra{r''}_j,
	\quad j\neq k
\eal
An external driving field coherently couples $\ket{g}$ with
$\ket{r}$, $\ket{r'}$ and $\ket{r''}$. When two atoms are in $\ket{r'r''}$ or
$\ket{r''r'}$ states, they interact via the Rydberg interaction. The
resulting Hamiltonian is
\bel
	\frac{1}{\hbar} H = \sum_j \frac{\Omega_0}{2} \big(\ket{G}\bra{j} +
	\text{h.c.}\big) + \sum_{j,k} \frac{\Omega_m}{2} \big(\ket{j}\bra{j,k} + \ket{j}\bra{k,j} +
	\text{h.c.}\big) +
	\sum_{j,k} \Delta_{jk} \ket{j,k}\bra{j,k}
\eel
In the ideal case, when $\Delta_{j,k}\rightarrow \infty$, $\ket{G}$ is
coherently coupled to the symmetric combination of a single excitation,
\bel
	\ket{S} = \frac{1}{\sqrt{N}}\sum_j \ket{j},
\eel
and the resulting dynamics is a Rabi oscillation between $\ket{G}$ and
$\ket{S}$, with Rabi frequency $\Omega_R = \sqrt{N} \Omega$, if the system
starts in $\ket{G}$.

To investigate the
deviation of the real dynamics from the ideal one, we focus on the coupling of
$\ket{j,k}$ states to $\ket{S}$. Using this notation the Hamiltonian can be
written as
\bel
\label{eq:H}
	\frac{1}{\hbar} H = \frac{\sqrt{N}\Omega_0}{2} \big(\ket{G}\bra{S} +
	\text{h.c.}\big) + \sum_{j,k>j} \frac{\Omega_m}{\sqrt{N}}
	\big(\ket{S}\bra{M_{jk}} + \text{h.c.}\big) + \sum_{j,k>j} \Delta_{jk}
	\ket{M_{jk}}\bra{M_{jk}},
\eel
where $\ket{M_{jk}} = \frac{\ket{j,k} + \ket{k,j}}{\sqrt{2}}$, and the
non-symmetric combinations are not coupled to $\ket{S}$.

\subsection{Broadening}

We adiabatically eliminate the doubly excited states $\{\ket{j,k}\}$ from
\refeq{eq:H} to arrive to the effective Hamiltonian,
\bel
\label{eq:Heff}	
	\frac{1}{\hbar} H_\text{eff} = h_{\Omega_0} + \sum_j \left(\Delta_j -
	i\frac{\Gamma_j}{2}\right)\ket{j}\bra{j} + \sum_{j,k>j}
	\frac{\Omega_{jk}}{2}\big(\ket{j}\bra{k} + \ket{k}\bra{j}\big),
\eel
where $h_{\Omega_0}$ is the first term in \refeq{eq:H}, and the new coefficients
are
\bal
	\Delta_j &=& -\sum_{k\neq j} \frac{\Omega_m^2}{4} \frac{\PP }{\Delta_{jk}},
	\\
	\Omega_{jk} &=& -\frac{\Omega_m^2}{2\Delta_{jk}},
	\\
	\Gamma_j &=& 2\pi \sum_{k\neq j}\Omega_m^2 \delta(\Delta_{jk}),
\eal
where $\PP$ indicates principal value. The $\Gamma_j$ terms describe the
resonant excitation, which is as discussed in the previous section.

The $j$ dependence of $\Delta_j$ results in inhomogeneous
broadening, $\delta\Delta := \sqrt{\ev{\Delta^2} - \ev{\Delta}^2}$.
\bel
	\delta\Delta =
	\frac{\Omega_m^2}{4}\sqrt{\text{Var}\left(\sum_{k\neq j
	}-\frac{\PP}{\Delta_{jk}}\right)} =
	\frac{\hbar \Omega_m^2}{4}\sqrt{\text{Var}\left(\sum_{k\neq
	j}\frac{\PP}{\Delta_E}\frac{R_{jk}^6}{R_{jk}^6 - R_\times^6}\right)},
\eel
where we used $\hbar \Delta_{jk} = U(R_{jk}) = \frac{C_6}{R_{jk}^6}-\Delta_E$, i.e. a van-der-Waals interaction potential between the Rydberg atoms, and eliminated $C_6$ by using that $R_\times^6=\frac{C_6}{\Delta_E}$.

The sum can be written as
\bal\label{eq:sumtoint}
	D_j := \sum_{k\neq j} \frac{\PP}{\Delta_E}\frac{R_{jk}^6}{R_{jk}^6 - R_\times^6}
	&=&
	\frac{N}{V\Delta_E}\intop_V d{^3\br_k}\frac{\PP |\br_k -\br_j|^6}{|\br_k -\br_j|^6 - R_\times^6}=
	\frac{N}{V\Delta_E}\intop_{V'} d{^3\br}\frac{\PP r^6}{r^6 - R_\times^6},
\eal
where the integrand is more conveniently written in terms of $r=|\br|=|\br_k-\br_j|$. $\br$ can be seen as `local' spherical coordinate, centered around $\br_j$. Due to global rotational invariance of the problem we can set $\br_j=r_j\hat{\mathbf{z}}$ without loss of generality. We then find (see appendix \ref{app:A})
\bel
	\frac{D_jV\Delta_E}{N} =\intop_0^{R-r_j}4\pi r^2 \frac{\PP r^6}{r^6 - R_\times^6}d{r}+\intop_{R-r_j}^{R+r_j}2\pi r^2\frac{\PP r^6}{r^6 - R_\times^6}\left(1-\frac{r^2 + r_j^2 -R^2}{2rr_j}\right)d{r}.
\eel
We now assume that the singularity is in the first integral, $R_\times<R-r_j$. As a result, the second integral is no longer a principal value integral and since $r\geq R-r_j>R_\times$ we will furthermore approximate $\frac{r^6}{r^6-R_\times^6}\approx1$. The second integral is now straightforward to evaluate. Finally we use the indefinite integral
\bal
	\int\frac{r^8}{r^6-R_\times^6}d{r}=\frac{1}{3}r^3-\frac{1}{3}R_\times^3\text{arctanh}\left(\frac{r^3}{R_\times^3}\right)
\eal
to find the remaining principal value integral. The result is
\bal
	D_j=\frac{N}{\Delta_E}\left[1-\left(\frac{R_\times}{R}\right)^3\text{arctanh}\left(\left(\frac{R_\times}{R-r_j}\right)^3\right)\right].
\eal
Now, we can determine the averages $\ev{D}$ and $\ev{D^2}$, but since the above expression for $D_j$ is only valid when $r_j<R-R_\times$ we modify the averages to only average over a sphere with radius $R-R_\times$:
\bal
	\ev{D} &=& \frac{1}{N}\sum_j D_j = \frac{1}{V}\intop_V\d{^3\br_j}D_j\approx \frac{3}{2(R-R_\times)^3}\intop_0^{R-R_\times}r_j^2D_jd{r_j},
	\\
	\ev{D^2} &=& \frac{1}{N}\sum_j (D_j)^2 =\frac{1}{V}\intop_V\d{^3\br_j}(D_j)^2\approx \frac{3}{2(R-R_\times)^3}\intop_0^{R-R_\times}r_j^2(D_j)^2d{r_j},\quad
\eal
where we also used that $D_j$ only depends on $r_j=|\br_j|$. Using the expansion
\bal
	D_j=\frac{N}{\Delta_E}\left(1-\frac{R_\times^6}{R^3(R-r_j)^3}+\ldots\right)
\eal
up to the first term including $r_j$ the averages are found to be
\bal
\ev{D} &=&\frac{N}{\Delta_E}\left[1-\frac{3}{2}\left(\frac{R_\times}{R}\right)^{4}+\frac{3}{2}\left(\frac{R_\times}{R}\right)^{5}+\ldots\right],\\
\ev{D^2} &=& \left(\frac{N}{\Delta_E}\right)^2\left[1-3\left(\frac{R_\times}{R}\right)^{4}+3\left(\frac{R_\times}{R}\right)^{5}+\ldots\right],
\eal
where we made an expansion in powers of $\frac{R_\times}{R}$. As a result we find (lowest order in $\frac{R_\times}{R}$)
\bel
	\delta D = \sqrt{\ev{D^2} - \ev{D}^2} =\frac{N}{\Delta_E}\sqrt{\frac{3}{5}}\left(\frac{R_\times}{R}\right)^{7/2}.
\eel
With this result we can write the broadening $\delta\Delta$ as
\bel
	\delta\Delta=\frac{\hbar\Omega_m^2}{4}\delta D=\sqrt{\frac{12}{5}}\pi\hbar\Omega_m^2\rho_d\frac{R^2}{|U'(R_\times)|}\left(\frac{R_\times}{R}\right)^{5/2},
\eel
where the power of $5/2$ was also reproduced using numerical calculations. If we apply the same calculations on a dipolar potential, $U(R_{jk})=\frac{C_3}{R_{jk}^3}-\Delta_E$, the result is
\bel
\delta\Delta=\frac{7}{2}\pi\hbar\Omega_m^2\rho_d\frac{R_\times^2}{|U'(R_\times)|},
\eel
which surprisingly has the same dependence on $R_\times$ as $\gamma_m$.

\subsection{Effect on coherence}
The effect of this inhomogeneous broadening is well approximated by an
additional decay of the coherence between the ground state $\ket{G}$ and the
symmetric single-excitation state $\ket{S} = \sum_i \ket{R_i}/N$, by a factor of
$e^{-(\delta\Delta\,t)^2/2}$. This is a much weaker effect than
the pure exponential decay, set by $\gamma_m$, and since $\delta\Delta \lesssim
\gamma_m$ for the parameter regime in consideration, the effect of inhomogeneous
broadening can be neglected as long as $\gamma_m t < 1$.

\section{Dephasing}

In the absence of molecular coupling, Rabi flopping occurs between the
collective ground state and the collective single Rydberg excitation%
\begin{align*}
|\tilde{R}^{I}\left(  t\right)  \rangle &  =\sum_{k=1}^{N}c_{k}^{I}\left(
t\right)  |R_{k}\rangle=i\sin\left(  \frac{\sqrt{N}\Omega_{0}t}{2}\right)
\frac{1}{\sqrt{N}}\sum_{k=1}^{N}|R_{k}\rangle\\
|\tilde{G}^{I}\left(  t\right)  \rangle &  =c_{g}^{I}\left(  t\right)
|G\rangle=\cos\left(  \frac{\sqrt{N}\Omega_{0}t}{2}\right)  |G\rangle\\
|\Psi^{I}\left(  t\right)  \rangle &  =|\tilde{G}^{I}\left(  t\right)
\rangle+|\tilde{R}^{I}\left(  t\right)  \rangle
\end{align*}
With the molecular resonances%
\begin{align*}
|\tilde{R}\left(  t\right)  \rangle &  =\sum_{k=1}^{N}\left(  c_{k}^{I}\left(
t\right)  +a_{k}\left(  t\right)  \right)  |R_{k}\rangle\\
|\tilde{G}\left(  t\right)  \rangle &  =\left(  c_{g}^{I}\left(  t\right)
+a_{g}\left(  t\right)  \right)  |G\rangle\\
|\Psi\left(  t\right)  \rangle &  =|\tilde{G}\left(  t\right)  \rangle
+|\tilde{R}\left(  t\right)  \rangle
\end{align*}
where we introduced corrections $a_{k}\left(  t\right)  $ to the ideal
amplitudes $c_{k}^{I}\left(  t\right)  $. In the ideal case

The overlap between the target $|\tilde{R}^{I}\left(  t\right)  \rangle$ and
the actual $|\tilde{R}\left(  t\right)  \rangle$ states reads
\begin{align*}
\langle\Psi^{I}\left(  t\right)  |\Psi\left(  t\right)  \rangle &  =\left(
c_{g}^{I}\left(  t\right)  \right)  ^{\ast}\left(  c_{g}^{I}\left(  t\right)
+a_{g}\left(  t\right)  \right)  +\sum_{k=1}^{N}\left(  c_{k}^{I}\left(
t\right)  \right)  ^{\ast}\left(  c_{k}^{I}\left(  t\right)  +a_{k}\left(
t\right)  \right)  =\\
&= \left(  c_{g}^{I}\left(  t\right)  \right)  ^{\ast}\left(
c_{g}^{I}\left(  t\right)  +a_{g}\left(  t\right)  \right)  +\sum_{k=1}%
^{N}\left\vert c_{k}^{I}\left(  t\right)  \right\vert ^{2}+\sum_{k=1}%
^{N}\left(  c_{k}^{I}\left(  t\right)  \right)  ^{\ast}a_{k}\left(  t\right)
=\\
&=  1+\left(  c_{g}^{I}\left(  t\right)  \right)  ^{\ast}a_{g}\left(  t\right)
+\sum_{k=1}^{N}\left(  c_{k}^{I}\left(  t\right)  \right)  ^{\ast}a_{k}\left(
t\right)
\end{align*}
Distance between $|\Psi^{I}\left(  t\right)  \rangle$ and $|\Psi\left(
t\right)  \rangle$%
\begin{align*}
d^{2} &  =\langle\Psi-\Psi^{I}|\Psi-\Psi^{I}\rangle=1+\langle\Psi\left(
t\right)  |\Psi\left(  t\right)  \rangle-2\operatorname{Re}\langle\Psi
^{I}\left(  t\right)  |\Psi\left(  t\right)  \rangle=\\
&  \left(  \langle\Psi\left(  t\right)  |\Psi\left(  t\right)  \rangle
-1\right)  -2\operatorname{Re}\left(  \left(  c_{g}^{I}\left(  t\right)
\right)  ^{\ast}a_{g}\left(  t\right)  +\sum_{k=1}^{N}\left(  c_{k}^{I}\left(
t\right)  \right)  ^{\ast}a_{k}\left(  t\right)  \right)
\end{align*}%
\[
\langle\Psi\left(  t\right)  |\Psi\left(  t\right)  \rangle
=1+2\operatorname{Re}\left(  \left(  c_{g}^{I}\left(  t\right)  \right)
^{\ast}a_{g}\left(  t\right)  +\sum_{k=1}^{N}\left(  c_{k}^{I}\left(
t\right)  \right)  ^{\ast}a_{k}\left(  t\right)  \right)  +\left\vert
a_{g}\right\vert ^{2}+\sum_{k}\left\vert a_{k}\right\vert ^{2}%
\]%
\[
d^{2}=\left\vert a_{g}\right\vert ^{2}+\sum_{k}\left\vert a_{k}\right\vert
^{2}%
\]

As in the derivation of the molecular exctition probability, consider a
transient coupling to a molecular state $|M_{ij}\rangle$. Consider time
evolution of the rydberg amplitude of atom $i$.
\begin{align*}
i\dot{c}_{i}  &  =\frac{1}{2}\Omega_{0}c_{g}+\frac{1}{2}\Omega_{m}m_{ij}\\
i\dot{c}_{g}  &  =\frac{1}{2}\Omega_{0}\sum_{i}^{N}c_{i}\,,\\
i\dot{m}_{ij}  &  =\Delta_{ij}m_{ij}+\frac{1}{2}\Omega_{m}^{ij}\left(
c_{i}+c_{j}\right)  .
\end{align*}
or with $c_{i}=c_{i}^{I}+a_{i},$ we find (here we take into account that the
time evolution of $a_{i}$ and $a_{j}$ are identical - this is true for the
first collision only)%
\begin{align*}
i\dot{a}_{g}  &  =\Omega_{0}a_{i}\\
i\dot{a}_{i}  &  =\frac{1}{2}\Omega_{0}a_{g}+\frac{1}{2}\Omega_{m}m_{ij}\\
i\dot{m}_{ij}  &  =\Delta_{ij}m_{ij}+\Omega_{m}c_{i}^{I}%
\end{align*}
as derived in Sec.%
\begin{align*}
m_{ij}\left(  t\right)   &  \approx m_{ij}\left(  t_{-}\right)  -i\Omega
_{m}c_{i}^{I}\left(  t\right)  \exp\left(  -i\phi\left(  t\right)  \right)
\int_{t_{-}}^{t}\exp\left(  i\phi\left(  t^{\prime}\right)  \right)
dt^{\prime}\\
&  m_{ij}\left(  t_{-}\right)  -i\Omega_{m}c_{i}^{I}\left(  t\right)
\exp\left(  -i\phi\left(  t\right)  \right)
\end{align*}
Since
\begin{align*}
i\frac{d^{2}}{dt^{2}}a_{i}  &  =\frac{1}{2}\Omega_{0}\frac{d}{dt}a_{g}%
+\frac{1}{2}\Omega_{m}\frac{d}{dt}m_{ij}\\
i\frac{d^{2}}{dt^{2}}a_{i}  &  =-i\frac{1}{2}\Omega_{0}^{2}a_{i}+\frac{1}%
{2}\Omega_{m}\frac{d}{dt}m_{ij}\\
\frac{d^{2}}{dt^{2}}a_{i}+\frac{1}{2}\Omega_{0}^{2}a_{i}  &  =-i\frac{1}%
{2}\Omega_{m}\frac{d}{dt}m_{ij}=-\frac{1}{2}\Omega_{m}\left(  \Delta
_{ij}m_{ij}+\Omega_{m}c_{i}^{I}\right)
\end{align*}
I want to neglect $\Delta_{ij}m_{ij}$. Let's do this first%
\[
\frac{d^{2}}{dt^{2}}a_{i}+\frac{1}{2}\Omega_{0}^{2}a_{i}=-\frac{1}{2}%
\Omega_{m}^{2}~\frac{1}{\sqrt{N}}\sin\left(  \frac{\sqrt{N}\Omega_{0}t}%
{2}\right)
\]
we can consider the two contributions separately as the solutions linearize in
the inhomogeneous part anyways. Basically these oscillations persist even
without the collisions (I guess one needs to look at time-dependence of
$\Omega_{m}$). The second part%
\[
\frac{d^{2}}{dt^{2}}a_{i}+\frac{1}{2}\Omega_{0}^{2}a_{i}=-\frac{1}{2}%
\Omega_{m}\Delta_{ij}m_{ij}%
\]

Actually let's leave the eqn in this form (to use integration by parts)%
\[
\frac{d^{2}}{dt^{2}}a_{i}+\frac{1}{2}\Omega_{0}^{2}a_{i}=-i\frac{1}{2}%
\Omega_{m}\frac{d}{dt}m_{ij}%
\]
Solution $\omega_{0}=\Omega_{0}/\sqrt{2}$%
\begin{align*}
a_{i}\left(  t_{x}+\delta t_{x}\right)   &  =b_{+}e^{i\omega_{0}t}%
+b_{-}e^{-i\omega_{0}t}+\\
&  -\frac{1}{4}\left(  \frac{\Omega_{m}}{\omega_{0}}\right)  ~e^{-i\omega
_{0}t_{+}}\int_{t_{x}-\delta t_{x}}^{t_{x}+\delta t_{x}}\exp\left(
+i\omega_{0}t^{\prime}\right)  \dot{m}\left(  t^{\prime}\right)  dt^{\prime}\\
&  +\frac{1}{4}\left(  \frac{\Omega_{m}}{\omega_{0}}\right)  ~e^{+i\omega
_{0}t_{+}}\int_{t_{x}-\delta t_{x}}^{t_{x}+\delta t_{x}}\exp\left(
-i\omega_{0}t^{\prime}\right)  \dot{m}\left(  t^{\prime}\right)  dt^{\prime}%
\end{align*}
The initial conditions are determined by
\[
a_{i}\left(  t_{x}-\delta t_{x}\right)  =b_{+}e^{i\omega_{0}t_{-}}%
+b_{-}e^{-i\omega_{0}t_{-}}%
\]%
\begin{align*}
\int_{t_{-}}^{t_{+}}\exp\left(  +i\omega_{0}t\right)  \dot{m}\left(
t^{\prime}\right)  dt^{\prime}  &  =\left.  \exp\left(  +i\omega_{0}t\right)
m\left(  t\right)  \right\vert _{t_{-}}^{t_{+}}-i\omega_{0}\int_{t_{x}-\delta
t_{x}}^{t_{x}+\delta t_{x}}\exp\left(  +i\omega_{0}t^{\prime}\right)  m\left(
t^{\prime}\right)  dt^{\prime}\\
\left.  \exp\left(  +i\omega_{0}t\right)  m\left(  t\right)  \right\vert
_{t_{-}}^{t_{+}}  &  =\\
-i\omega_{0}\int_{t_{x}-\delta t_{x}}^{t_{x}+\delta t_{x}}\exp\left(
+i\omega_{0}t^{\prime}\right)  m\left(  t^{\prime}\right)  dt^{\prime}  &
=-i\omega_{0}m_{ij}\left(  t_{-}\right)  \int_{t_{x}-\delta t_{x}}%
^{t_{x}+\delta t_{x}}\exp\left(  +i\omega_{0}t^{\prime}\right)  dt^{\prime}+\\
&  -\left(  -i\omega_{0}\right)  i\Omega_{m}c_{i}^{I}\left(  t\right)
\int_{t_{x}-\delta t_{x}}^{t_{x}+\delta t_{x}}\exp\left(  +i\omega
_{0}t\right)  \exp\left(  -i\phi\left(  t\right)  \right)  dt\int_{t_{-}}%
^{t}\exp\left(  i\phi\left(  t^{\prime}\right)  \right)  dt^{\prime}%
\end{align*}

\section{Effect of attractive potential}
The attractive potentials (see at
$R_\times \approx
4.6 \, \mu\mathrm{m}$ in Fig.~1 of the paper) can lead to
auto-ionization in the small $R$ region~\cite{Hahn2000}. Such a process  would
free an electron and a molecular ion, with their Coulomb fields  blockading the
entire sample. While in our illustrative example, laser coupling to attractive
potentials is negligible, it may be not be the case in general. Qualitatively,
to reduce auto-ionization one needs to pick Rydberg states such that the
potentials inside the most strongly coupled resonance shells are repulsive.

\newpage
\section{Suppression of resonance by trapping in regular lattice}
We investigate the possibility of trapping the atoms in an 3D cubic optical
lattice, in order to suppress the effective decay rate $\gamma$, due to
double Rydberg resonances. We use the homogeneous atom density case as a
benchmark,
\bel
	\gamma_\text{hom} = \sum_j \gamma_j = 2\pi^2 \Omega_0  \rho
	\sum_i\xi_i R_{\times,i}^2 \Delta R_{\times,i},
\eel
where the summation is performed over all resonances at the crossing distances
$R_{\times,i}$, each having a width of $\Delta R_{\times,i}$ and molecular
coupling factor $\xi_i$. Here $\rho$ is assumed to be constant.

We assume that in a 3D cubic lattice, each lattice site holds a single atom,
trapped in the ground state of the harmonic trap. Let $a$ denote the lattice
constant and $d$ be the size of each trapped wavefunction. The deeper we make
the lattice, the smaller the $d/a$ ratio can be. For given $d, a$ values, we can
plot the 3D density $\rho(R)$ of the atoms as a function of distance from a
particular lattice site. This is shown on \reffig{fig:rho_3D} with a blue curve.
\begin{figure}[h]
\centering
\includegraphics[width=0.85\textwidth]{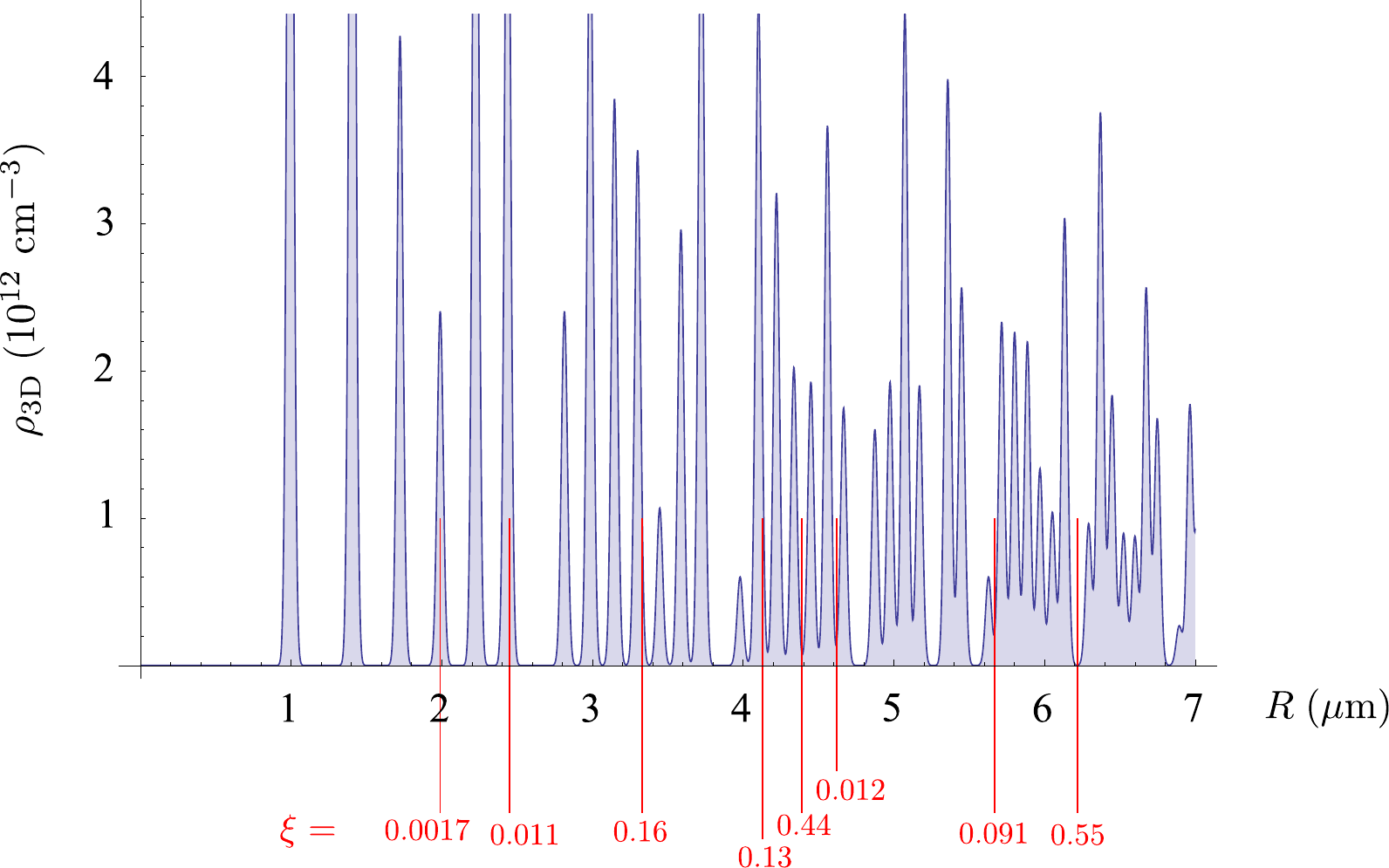}
\caption{
\label{fig:rho_3D}
3D density of the atoms in a cubic lattice (lattice constant: $a =
0.995~\mu\text{m}$) as a function of distance, $R$. Each atom is confined by a
harmonic trap to a region of size $d = 0.02~\mu\text{m}$. Red vertical lines
indicate the position of the resonances given in Table 1 of the paper. The
numbers shown next under the lines are the moleculare coupling coefficients,
$\xi_i$, for each resonance.}
\end{figure}

By taking the $R$ dependence of the atom density $\rho(R)$ into account, we can
write the total decay rate as
\bel
	\gamma_\text{lattice} = \sum_j \gamma_j = 2\pi^2 \Omega_0
	\sum_i \rho(R_{\times,i}) \xi_i R_{\times,i}^2 \Delta R_{\times,i},
\eel
numerically evaluate, and compare it with the homogeneous result,
$\gamma_\text{hom}$. On \reffig{fig:gamma_plot}, we plot $\gamma_\text{lattice}
/ \gamma_\text{hom}$ as a function of the trap confinement $d$ for different
fixed values of $a$. A confinement of $d/a \approx 0.01$ can suppress the decay
to 2--10\% of its homogeneous density value, depending on the accuracy of the
fine tuning of the lattice constant.
\begin{figure}[h]
\centering
\includegraphics[width=0.6\textwidth]{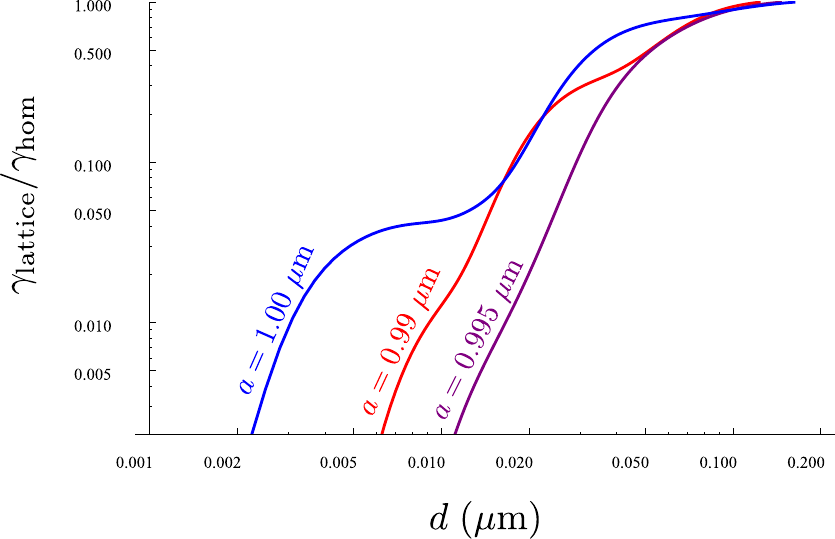}
\caption{
\label{fig:gamma_plot}
Decay rate in a 3D lattice normalized with the homogeneous decay rate result
for lattice constants $a = 1.00,\,0.995$ and $0.99~\mu\text{m}$.
The suppression is strong for small trap size, and diminishes
($\gamma_\text{lattice} \rightarrow \gamma_\text{hom}$) for large size.  Small
changes in $a$ result in significant changes in the suppression. This is due to
the detailed peak structure of the 3D density in the lattice. As a result, fine
tuning of the lattice constant is required.}
\end{figure}

\newpage
\begin{appendix}
\section{Local spherical coordinates}
\label{app:A}
Since we make the substitution $\br=\br_k-\br_j$ the boundaries of the polar integral change. Depending on the relation between $r$, $r_j$ and $R$ there are three regions, as shown on \reffig{fig:spheres}.
First, if $r < R-r_j$, then the entire sphere of radius $r=|\br|$ lies within the
boundaries of the cloud, and therefore we have $0\leq\theta\leq\pi$. Similarly, if $r > R
+ r_j$, then the opposite is true: the cloud lies entirely inside the sphere of
radius $r$, and therefore there is no contribution to the integral from this part.

\begin{figure}[h]
\centering
\includegraphics[width=0.9\textwidth]{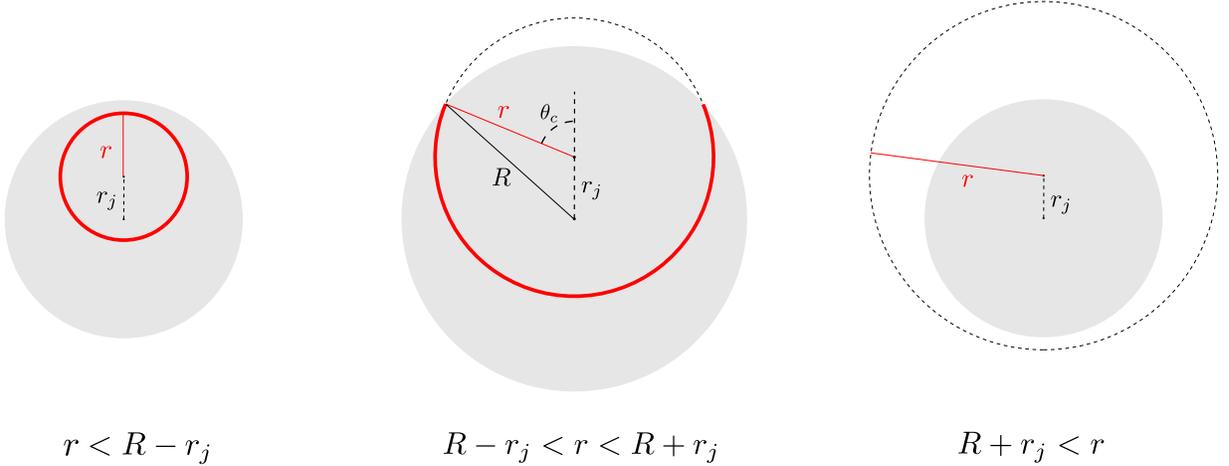}
\caption{
\label{fig:spheres}
We only need to integrate over that part (red) of the dotted sphere that is inside the cloud. This corresponds to a modification of the boundaries for the $\theta$-integral, depending on the relation between $r$, $r_j$ and $R$.}
\end{figure}
% Figure still needs to be adapted!

If $R-r_j < r < R+r_j$, then there exist a circle, where the sphere of radius $r$
intersects the boundary of the cloud, as shown on \reffig{fig:spheres}. The
angle between the segments $r_j$ and $r$ is
\bel
	\pi-\theta_c = \arccos\left(\frac{r^2 + r_j^2 -R^2}{2rr_j}\right),
\eel
and therefore we have $\theta_c\leq\theta\leq\pi$. Thus we can rewrite the integral in \refeq{eq:sumtoint} as
\bel
	\intop_{V'}\d{^3\br}f(r)=\intop_0^{R-r_j}r^2d{r}\intop_0^{\pi}\sin(\theta)d{\theta}\intop_0^{2\pi}d{\phi} f(r)+\intop_{R-r_j}^{R+r_j}r^2d{r}\intop_{\theta_c}^{\pi}\sin(\theta)d{\theta}\intop_0^{2\pi}d{\phi}f(r)
\eel
where $f(r)$ is the integrand that only depends on $r$. Evaluating the integrals over $\phi$ and $\theta$ we find
\bal
\label{eq:splitted}
	\intop_{V'}\d{^3\br}f(r)=\intop_0^{R-r_j}4\pi r^2 f(r)d{r}+\intop_{R-r_j}^{R+r_j}2\pi r^2f(r)\left(1-\frac{r^2 + r_j^2 -R^2}{2rr_j}\right)\d{r}.
\eal
\end{appendix}
\end{widetext}

%\bibliographystyle{../prl2012}
%\bibliography{../library}

\end{document}